# 21-Component Compositionally Complex Ceramics: Discovery of Ultrahigh-Entropy Weberite and Fergusonite Phases and a Pyrochlore-Weberite Transition

Mingde QIN [a], Heidy VEGA [a,b], Dawei ZHANG [a,c], Sarath ADAPA [d], Andrew J. WRIGHT [a], Renkun CHEN [c,d], Jian LUO [a,c,*]

[a] Department of NanoEngineering; [b] Department of Chemistry and Biochemistry; [c] Program of Materials Science and Engineering; [d] Department of Mechanical and Aerospace Engineering,

University of California San Diego, La Jolla, CA 92093, USA

**Abstract**

Two new high-entropy ceramics (HECs) in the weberite and fergusonite structures, along with unexpected formation of ordered pyrochlore phases with ultrahigh-entropy compositions and an abrupt pyrochlore-weberite transition, are discovered in a 21-component oxide system. While the Gibbs phase rule allows 21 equilibrium phases, nine out of the 13 compositions examined possess single HEC phases (with ultrahigh ideal configurational entropies: ~$2.7k_B$ per cation or higher on one sublattice in most cases). Notably, $(15RE_{1/15})(Nb_{1/2}Ta_{1/2})O_4$ possess a single monoclinic fergusonite (C2/c) phase and $(15RE_{1/15})_3(Nb_{1/2}Ta_{1/2})_1O_7$ form a single orthorhombic (C2221) weberite phase, where $15RE_{1/15}$ represents $Sc_{1/15}Y_{1/15}La_{1/15}Pr_{1/15}Nd_{1/15}Sm_{1/15}Eu_{1/15}Gd_{1/15}Tb_{1/15}Dy_{1/15}Ho_{1/15}Er_{1/15}Tm_{1/15}Yb_{1/15}Lu_{1/15}$. Moreover, a series of eight $(15RE_{1/15})_{2+x}(Ti_{1/4}Zr_{1/4}Ce_{1/4}Hf_{1/4})_{2-2x}(Nb_{1/2}Ta_{1/2})_xO_7$ specimens all exhibit single phases, where a pyrochlore-weberite transition occurs within $0.75 < x < 0.8125$. This cubic-to-orthorhombic transition does not change the temperature-dependent thermal conductivity appreciably, as the amorphous limit may have already been achieved in the ultrahigh-entropy 21-component oxides. These discoveries expand the diversity and complexity of HECs, towards many-component compositionally complex ceramics (CCCs) and ultrahigh-entropy ceramics.

---

[*] Correspondence should be address to J.L. (email: jluo@alum.mit.edu)

# 1. Introduction

High-entropy ceramics (HECs) [1-3], which can be generalized to a broader class of compositionally complex ceramics (CCCs) [1], are emergent as a new research field (Fig. 1). In the last few years, single-phase high-entropy oxides [4-7], borides [8-14], silicides [15-17], carbides [18-24], nitrides and carbonitrides [25, 26], fluorides [27, 28], sulfides [29], and intermetallic aluminides [30] have been fabricated. Among HECs, high-entropy oxides in the fluorite [1, 6, 31-36], rocksalt [4], pervoskite [5, 37, 38], pyrochlore [39-41], and spinel [42] crystal structures, as well as more complex silicates [43], phosphates [44], aluminates [45, 46], and molybates [47], have been studied. The vast majority of prior HEC studies focused on five-component equimolar compositions that produce ~1.61$k_B$ per cation ideal configurational entropy on at least one sublattice, where $k_B$ is the Boltzmann constant, but some studies also include four-component [20] and six- to nine-component [48] equimolar compositions [1-3]. Recent studies further investigated the high-entropy phase formation and transition in non-equimolar 10- and 11-component oxide systems, which can form either disordered (fluorite) or ordered (pyrochlore) phase (with either one or two cation sublattices) [31, 49]. This study aims at further exploring many-component CCCs and discovering new high-entropy (including ultrahigh-entropy) phases; see relevant definitions of "many-component CCCs" and "ultrahigh-entropy ceramics" in Fig. 1 and subsequent discussions. Specifically, this study reported nine new single-phase ultrahigh-entropy compositions in three crystal structures, each with 17 to 21 different metal cations. Notably, this study discovered two new (weberite and fergusonite) HECs, somewhat unexpected formation of ordered pyrochlore phases with ultrahigh-entropy compositions, and a rather abrupt pyrochlore-weberite transition in a 21-component oxide system.

Akin to their metallic counterparts [50], we can propose to expand HECs to a broader class of CCCs or "compositionally complex ceramics", as schematically illustrated in Fig. 1 [1]. The more generalized CCCs can further include or consider medium-entropy compositions, non-equimolar compositional designs, and/or short- and long-range cation ordering, which often reduce the configurational entropy, but offer additional complexity, more tunability, and potentially new or improved properties [1, 17, 31, 36, 39, 51]. On the one hand, the configurational entropy can sometimes be reduced to less than 1.5$k_B$ per cation (a somewhat subjective threshold to define HECs [1]) because of cation ordering in a single-phase CCC [17] or non-equimolar cation partitions (dictated by a thermodynamic equilibrium) in a dual-phase CCC [51], even for overall equimolar five-component compositions. On the other hand, we can further explore multiple cation sublattices, different crystal structures, mixed ionic-covalent-metallic bonding, and/or defects (*e.g.*, aliovalent doping and oxygen vacancies) to embrace and exploit the complexity.

While the vast majority of prior studies investigated four- [20] to nine-component [48] (mostly equimolar) compositions [1-3], we propose to further explore (equimolar and non-equimolar) "many-component CCCs" (*i.e.*, ≥10 components), which can offer even larger (and tailorable) compositional spaces and further (extreme) complexity (Fig. 1). These many-component CCCs include ultrahigh-entropy ceramics (a subclass of HECs), which we define as those HECs with >2.3$k_B$ per cation ideal configurational entropy on at least one sublattice; noting that ln(10) ≈ 2.3, so that equimolar many-component CCCs with ≥10 disordered components should be ultrahigh-entropy. However, non-equimolar compositions and cation ordering can reduce the configurational entropy of many-component CCCs so that they (may or) may not be ultrahigh-entropy (or even high-entropy) ceramics (just like CCCs *vs.* HECs), as shown in Fig. 1. We note that these definition thresholds adopted here (10 components and 2.3$k_B$ per cation) are somewhat arbitrary and subjective (just like five components and 1.5$k_B$ per cation thresholds commonly used for HECs [1] and their metallic counterparts, high-entropy alloys [50]). Here, we recognize (and emphasize) that maximizing the entropy may not be necessary (or beneficial) in many cases; instead, the compositional and structural complexities offer new opportunities to tailor the phase stability and properties of CCCs [1, 17, 31, 36, 39, 51]. In general, the complexities in (many-component) CCCs offer more opportunities, where different or new phenomena can emergent in complex systems, because "more is different" (P. W. Anderson, 1972 [52]). For example, the somewhat unexpected properties of the 21-component CCCs observed in this study include the unusual small numbers of (only one or two) equilibrium phases formed, unpredicted formation of ordered pyrochlore phases (*vs.* the predicted disordered defect-fluorite) with ultrahigh-entropy compositions, and a surprisingly abrupt transition between two phases of identical compositions but different crystal structures within a narrow compositional region.

Compositionally complex fluorite-based oxides provide a platform to investigate the phase stability of various derivative "ordered" and often distorted (superlattice) structures via changing the ratios of cations of different valences and radii. Here, cubic $AO_2$ fluorite ($Fm\bar{3}m$, No. 225) of a disordered cation arrangement (with only one cation sublattice A) is the primitive structure, while cubic $A_2B_2O_7$ pyrochlore ($Fd\bar{3}m$, No. 227), orthorhombic $A_3BO_7$ weberite ($C222_1$, No. 20), and monoclinic $ABO_4$ fergusonite ($C2/c$, No. 15) can be considered as derivative structures with oxygen vacancies, cation ordering, and lattice distortion (with two cation sublattices A and B), as shown in Fig. 2. Here, we use "weberite" and "fergusonite" (or, more rigorously, "weberite-type" and "fergusonite-type" structures,) to only denote the aforementioned space groups, while we note that both of them can refer to multiple polymorphs (of different space groups and specific crystal structures) in mineralogy [53, 54]. Recent studies uncovered long-range order-disorder (pyrochlore-fluorite) transitions in non-equimolar 10- and 11-component HECs/CCCs [31, 49] and revealed short-range weberite ordering (at the nanoscale) in long-range defect-fluorite structured high-entropy rare earth (RE) niobates/tantalates $RE_3(Nb/Ta)O_7$ [32].

In this study, we investigate a 21-component Nb-Ta-Ti-Zr-Ce-Hf-Sc-Y-La-Pr-Nd-Sm-Eu-Gd-Tb-Dy-Ho-Er-Tm-Yb-Lu-O system. While the Gibbs phase rule allows the co-existence of up to 21 oxide phases in this 21-component oxide system (assuming fixed $P_{O2}$, temperature, and pressure), nine out of the 13 compositions examined possess single high-entropy (in fact, ultrahigh-entropy) phases in three fluorite-derived structures (pyrochlore, weberite, and fergusonite) and four others are dual-phase HECs. Notably, weberite and fergusonite HECs have never been reported previously. This study also discovered somewhat unexpected formation of the ordered pyrochlore phases (*vs.* the disordered defect-fluorite) with the ultrahigh-entropy compositions, along with an unexpectedly abrupt pyrochlore-weberite transition, in the 21-component ultrahigh-entropy oxide system.

## 2. Experimental

As shown in Fig. 2, we designed a compositional triangle with the following three endmembers: $(15RE_{1/15})_2(Ti_{1/4}Zr_{1/4}Ce_{1/4}Hf_{1/4})_2O_7$, $(15RE_{1/15})_3(Nb_{1/2}Ta_{1/2})_1O_7$, and $(15RE_{1/15})(Nb_{1/2}Ta_{1/2})O_4$ (also written as $(15RE_{1/15})_2(Nb_{1/2}Ta_{1/2})_2O_8$ to have the same number of sites per cell/formula with the other two). Here, $15RE_{1/15}$ represents $Sc_{1/15}Y_{1/15}La_{1/15}Pr_{1/15}Nd_{1/15}Sm_{1/15}Eu_{1/15}Gd_{1/15}Tb_{1/15}Dy_{1/15}Ho_{1/15}Er_{1/15}Tm_{1/15}Yb_{1/15}Lu_{1/15}$. Our experiments showed that the three endmembers possess single-phase pyrochlore, weberite, and monoclinic fergusonite structures, respectively. Thus, they are denoted as P0, W0, and MF, respectively. Here, we refer the monoclinic fergusonite endmember as "MF" to avoid confusion with the base fluorite structure. We use P0 and W0 to denote the 19-component pyrochlore and 17-component weberite endmembers, respectively, as we also found six additional 21-component compositions that are in single-phase pyrochlore and weberite structures along the P0-W0 line, which are denoted as P1, P2, P3, W3, W2, and W1, respectively.

As shown in Supplementary Fig. S1, the P0-W0-MF compositional triangle is a section of the larger $(15RE_{1/15})_2O_3$-$(Ti_{1/4}Zr_{1/4}Ce_{1/4}Hf_{1/4})O_2$-$(Nb_{1/2}Ta_{1/2})_2O_5$ compositional triangle. Thus, we first prepared three mixtures composing $(15RE_{1/15})_2O_3$, $(Ti_{1/4}Zr_{1/4}Ce_{1/4}Hf_{1/4})O_2$, and $(Nb_{1/2}Ta_{1/2})_2O_5$, respectively, by weighting out appropriate amounts of commercial binary oxides (~5 μm particle sizes, US Research Nanomaterials, TX, USA). A batch of 15 g was desired for each mixture and was placed in a 100 ml $Y_2O_3$-stablized $ZrO_2$ (YSZ) planetary milling jar with 10 ml of isopropyl alcohol and YSZ milling media at a ball-to-powder mass ratio of 10:1. The planetary ball mill was conducted at 300 RPM for 24 h in a PQN04 mill (Across International LLC, NJ, USA) to mix and homogenize the powders, and the mixed powders were dried at 75 °C overnight. The dried powders were ground into fine particles with an agate mortar and pestle. Subsequently, these three powders were used as precursors to fabricate specimens of the 13 different compositions illustrated in Fig. 2.

To fabricate each specimen, appropriate amounts of three aforementioned precursors were weighted out in batches of 2.5 g based on the compositions labeled in Fig. 3, Fig. 4 and Supplementary Fig. S2-S14. The mixed powders were then placed in a poly(methyl methacrylate) high-energy ball mill (HEBM) vial with tungsten carbide (WC) inserts and one Ø5/16" (~7.94 mm) WC ball. Another 1 wt.% (0.025 g) stearic acid was added as lubricant. The HEBM was conducted in a SPEX 8000D mill (SPEX CertiPrep, NJ, USA) for 100 min, and the as-milled powders were then pressed into green pellets at 100 MPa in a Ø1/2" (12.7 mm) stainless steel die for 2 min. After that, the green pellets were placed on a Pt foil in an alumina combustion boat and sintered inside a muffle box furnace (SentroTech, OH, USA) at 1600 °C for 24 h with 5 °C/min heating rate and furnace cooling. The sintered pellets were successively ground and polished before further characterizations.

A Rigaku Miniflex II diffractometer was utilized to collect the X-ray diffraction (XRD) data at 30 kV and 15 mA. For each single-phase specimen, unit cell refinement was performed through JADE to obtain the lattice parameters, and the theoretical density was calculated based on measured lattice parameters and the composition. Scanning electron microscopy (SEM) and energy dispersive X-ray spectroscopy (EDS) were conducted on a Thermo-Fisher (formerly FEI) Apreo microscope equipped with Oxford N-Max$^N$ EDS detector to examine the microstructure and compositional homogeneity. Scanning transmission electron microscopy (STEM) specimens were prepared by a Thermo-Fisher Scios focused ion beam (FIB), and STEM and nanoscale EDS were conducted at 300 kV on an JEOL JEM-ARM300CF aberration-corrected microscope (AC-STEM) with a high-angle annular dark-field (HAADF) detector.

Bulk densities were measured abiding by ASTM Standard C373-18 utilizing boiling water. Young's modulus ($E$) was determined following ASTM Standard C1198-20 using a Tektronix TDS 420A digital oscilloscope with 20 MHz longitudinal ultrasonic wave and 5 MHz transverse ultrasonic wave. Measured modulus was corrected for porosity according to $E = E_{measured}/(1 - 1.29P)$ [55], where $P$ is the pore fraction. Thermal conductivity ($k$) was calculated as the product of thermal diffusivity, bulk density, and heat capacity. Specimens were first coated with black carbon to maximize laser absorption and infrared emission. The thermal diffusivity was measured by a laser flash analyzer (LFA 467 HT HyperFlash, NETZSCH) and fit with a transparent model with the consideration of the radiative heat transfer (Supplementary Method and Fig. S16 for the detailed procedure and discussion). The heat capacity was calculated by the Neumann-Kopp rule using the data from Ref. [56]. Similar to Young's modulus, thermal conductivity was also corrected for porosity using the formula: $k = k_{measured}/(1 - P)^{3/2}$ [57]. Thus, all Young's modulus and thermal conductivity values reported in this study represent the intrinsic properties for fully dense materials (with the porosity effects removed). For the high-temperature data, the reported $k$ values represent the true (net) thermal conductivity values after removing the radiative heat transfer contributions (Supplementary Method).

## 3. Results and discussion

Fig. 2 illustrates the P0-W0-MF compositional triangle and the 13 specimens investigated in this study. All the three endmembers (P0, W0, and MF) are single-phase HECs that form three different fluorite-based crystal structures. First, endmember P0, $(15RE_{1/15})_2(Ti_{1/4}Zr_{1/4}Ce_{1/4}Hf_{1/4})_2O_7$, is in the cubic $A_2B_2O_7$ pyrochlore structure ($Fd\bar{3}m$, No. 227). Second, endmember W0, $(15RE_{1/15})_3(Nb_{1/2}Ta_{1/2})_1O_7$, is in the $Y_3TaO_7$-prototyped orthorhombic $A_3BO_7$ weberite-type structure ($C2221$, No. 20). Third, endmember MF, $(15RE_{1/15})_2(Nb_{1/2}Ta_{1/2})_2O_8$ (with the actual unit cell and irreducible formula $(15RE_{1/15})(Nb_{1/2}Ta_{1/2})O_4$), is the $YNbO_4$-prototyped $ABO_4$ monoclinic fergusonite-type structure ($C2/c$, No. 15). Here, we adopt the unique $c$-axis for the monoclinic fergusonite, which can alternatively be interpreted as an $I2/a$ structure with the unique $b$-axis. A combination of XRD characterization and EDS elemental mapping confirmed that these three endmembers are compositionally homogenous without detectable secondary phase. See Supplementary Figs. S2, S9, and S10 for the detailed XRD and EDS characterization results case by case, and the XRD patterns are also shown in Fig. 3 and Fig. 4. Since lower crystal symmetry will result in more diffraction peaks, the orthorhombic weberite and monoclinic fergusonite structures possess more diffraction peaks in the corresponding XRD patterns than the cubic pyrochlore.

To the best of our knowledge, HECs of the weberite-type and fergusonite-type structures have not been reported to date. Thus, our 17-component W0 $(15RE_{1/15})_3(Nb_{1/2}Ta_{1/2})_1O_7$ and MF $(15RE_{1/15})_2(Nb_{1/2}Ta_{1/2})_2O_8$ represent the two new HEC phases discovered. They both possess ultrahigh ideal configurational entropy of ~2.71$k_B$ per cation on the A sublattice and the overall mean configurational entropies of ~2.2$k_B$ per cation for W0 and ~1.7$k_B$ per cation for MF, averaged for A and B sublattices (Supplementary Table S2). Thus, they can be classified as ultrahigh-entropy ceramics (a subclass of HECs with >2.3$k_B$ per cation ideal configurational entropy on at least one sublattice, as well as a subclass of many-component CCCs, as schematically illustrated in Fig. 1). Here, we have ignored the anti-site disorder (entropy-driven swapping of A vs. B site cations), which is inevitable and can increase the entropy further, in calculating these ideal configurational entropies.

It was suggested that orthorhombic weberite (instead of the defect-fluorite) should form for stoichiometry $A_3B_1O_7$ with a larger ratio of cation radii of $r_{A^{3+}}/r_{B^{5+}} > $ ~1.40 based on the data for ternary niobates and tantalates [53, 58-61]. Our 17-component W0 $(15RE_{1/15})_3(Nb_{1/2}Ta_{1/2})_1O_7$ has a ratio of average cation radii $\overline{r_{A^{3+}}}/\overline{r_{B^{5+}}} \approx 1.56$ (calculated based on the ionic radii from Ref. [62]), which is consistent the observed weberite-type phase. However, we also note that this criterion for forming weberite vs. fluorite structure is not always held for ternary oxides. In fact, $Y_3NbO_7$ was known to form the defect-fluorite structure [63] and $Y_3TaO_7$ was shown to be stable in both structures [63-65], even both of them have similar high $r_{A^{3+}}/r_{B^{5+}}$ values.

The fergusonite structure has been widely reported in ternary RE niobates ($RENbO_4$) and tantalates ($RETaO_4$) [66]. Although HECs of the fergusonite-type structure have not been fabricated, natural $YNbO_3$-based minerals of the fergusonite-type structure can contain more than 10 cations (with Y and Nb being the major metal cations, so they are many-component CCCs, but may not be "high-entropy" yet) [67]. Thus, the 17-component MF $(15RE_{1/15})(Nb_{1/2}Ta_{1/2})O_4$ represents the first HEC fabricated in the monoclinic fergusonite-type structure (and they are also ultrahigh-entropy ceramics, a subgroup of HECs). While most prior studies of HECs have been focused on high-symmetry (cubic, hexagonal, and tetragonal) phases, the studies of low-symmetry orthorhombic and monoclinic HECs are less common (but with several exceptions, *e.g.*, high-entropy orthorhombic pervoskite [38], silicide carbides [68], monoborides [9], and $M_3B_4$ borides [10], as well as monoclinic silicates [43, 69] and phosphates [44]).

Although high-entropy pyrochlore oxides have been reported in several prior studies [39-41], the discovery of P0 $(15RE_{1/15})_2(Ti_{1/4}Zr_{1/4}Ce_{1/4}Hf_{1/4})_2O_7$ is interesting not only because it is ultrahigh-entropy ($\sim 2.71 k_B$ per cation on the A sublattice and $\sim 2.05 k_B$ per cation for the overall mean on both sublattices; Supplementary Table S2) but also because its formation is somewhat unexpected based on the criterion for ternary oxides or conventional high-entropy pyrochlore oxides. Here, it is well established based on ternary oxides that stoichiometry $A_2B_2O_7$ should form ordered pyrochlore structure with a larger ratio of cation radii $r_{A^{3+}}/r_{B^{4+}} > \sim 1.46$, while the (disordered) defect fluorite structure should form for a smaller ratio [70]. This criterion has been extended and verified for conventional high-entropy compositions based on the average cation radii with the same pyrochlore-fluorite transition threshold of $\overline{r_{A^{3+}}}/\overline{r_{B^{4+}}} \sim 1.46$ [71]. That study [71] proposed that a large size disorder [39] $\delta^* > \sim 5\%$ may lead to a dual-phase region near the threshold for $1.4 < \overline{r_{A^{3+}}}/\overline{r_{B^{4+}}} < 1.5$). Notably, our 19-component P0 $(15RE_{1/15})_2(Ti_{1/4}Zr_{1/4}Ce_{1/4}Hf_{1/4})_2O_7$ has a ratio of average cation radii $\overline{r_{A^{3+}}}/\overline{r_{B^{4+}}} \approx 1.43$ (below the threshold of $\sim 1.46$) and exceptionally high size disorder: $\delta_A \approx 6.6\%$, $\delta_B \approx 13\%$, and $\delta^* \equiv \sqrt{\delta_A^2 + \delta_B^2} \approx 14.6\%$ for the ordered pyrochlore. The size disorder would be even higher ($\delta_{disorder} = 19.9\%$) if a disordered defect-fluorite structure formed, where we used the same cation radii in the ordered pyrochlore (CN = 8 for the A-site RE elements and CN = 6 for the B-site 4+ cations based on Ref. [62]) for simplicity and a better comparison. However, we observed a single ultrahigh-entropy pyrochlore phase. This result suggests a deviation of the pyrochlore *vs.* fluorite stability rules (that would predict defect-fluorite based on ternary oxides or dual-phase) for this ultrahigh-entropy composition. As we will show subsequently, it is even more surprising that this ultrahigh-entropy pyrochlore structure is highly stable even after adding significant amounts of $Nb^{5+}$ and $Ta^{5+}$ cations to change the A:B stoichiometry to an equivalent *y* value of 0.75 (in our specimen P3), while ternary $A_{2\pm y}B_{2\mp y}O_{7\mp y/2}$ can disorder (to form the defect fluorite structure) at $y > \sim 0.1$-$0.3$ [31, 72].

To investigate pyrochlore *vs.* weberite phase stability and transformation, we further prepared and analyzed six specimens along the P0-W0 edge in Fig. 2 between the P0 and W0 endmembers, with the compositions $(15RE_{1/15})_{2+x}(Ti_{1/4}Zr_{1/4}Ce_{1/4}Hf_{1/4})_{2-2x}(Nb_{1/2}Ta_{1/2})_xO_7$ ($x$ = 0.25, 0.5, 0.75, 0.8125, 0.875, and 0.9375). The results are summarized in Supplementary Table S1 and documented Supplementary Figs. S2-S9. Interestingly, all the six 21-component ultrahigh-entropy compositions form single-phase HECs in either pyrochlore or weberite structure (Fig. 3 and Supplementary Fig. S3-S8). Thus, we named them as P1 ($x$ = 0.25), P2 ($x$ = 0.5), and P3 ($x$ = 0.75) for the three 21-component pyrochlore oxides and W3 ($x$ = 0.8125), W2 ($x$ = 0.875), and W1 ($x$ = 0.9375) for the three 21-component weberite oxides. Moreover, a rather abrupt pyrochlore-to-weberite transition (PWT) occurs within a narrow compositional range between $x$ = 0.75 and 0.8125, where the $(222)_p$ peak starts to split into $(202)_w$ and $(220)_w$ peaks due to the cubic-orthorhombic transition ($b_w \neq c_w$ in orthorhombic weberite), as shown in Fig. 3(b). Due to the compositional and structural complexities, it is infeasible to determine the distributions for each of 21 cations at A *vs.* B sites. Thus, we cannot calculate the exact configurational entropies for this series of "mixed" compositions (with non-stoichiometric ratios of A-type *vs.* B-type cations). However, we can estimate the configurational entropies for ideal cases with random (high-entropy) or preferential (low-entropy) anti-site mixing, as shown for P3 and W3 in Supplementary Table S2, which are generally higher than the endmembers. In brief, the long-range pyrochlore *vs.* weberite ordering with ultrahigh-entropy compositions, as well as an abrupt PWT or cubic-to-orthorhombic transition, is unequivocally evident (but it is uncertain it is a first-order transition).

Given the Gibbs phase rule that allows the co-existence of up to 21 equilibrium phases, the single-phase formation with a quite abrupt transition along the P0-W0 line is surprising and scientifically interesting. Here, we do not infer that the PWT or cubic-to-orthorhombic transition in 21-component oxides with change *x* must be a rigorous first-order transition defined in thermodynamics (as we feel that it is probably unlikely); but it does occur within a narrow compositional region of $\Delta x$ = 0.0625 (that we cannot resolve further experimentally, given the complexity of the system and the already subtle differences in the splitting of the XRD peaks for $x$ = 0.75 *vs.* 0.8125, as shown in Fig. 3). We should note that it is also surprising that the ultrahigh-entropy pyrochlore structure (with a $\overline{r_{A^{3+}}}/\overline{r_{B^{4+}}}$ ratio of ~1.43 in P0, lower than the normal threshold of ~1.46 to disorder) is stable in the ordered pyrochlore structure (against forming disordered defect-fluorite structure, either as a secondary phase or via an order-disorder transition overall) even after adding significant amounts of $Nb^{5+}$ and $Ta^{5+}$ cations in the series of $(15RE_{1/15})_{2+x}(Ti_{1/4}Zr_{1/4}Ce_{1/4}Hf_{1/4})_{2-2x}(Nb_{1/2}Ta_{1/2})_xO_7$. This addition not only introduces aliovalent doping, but also changes the A:B stoichiometry. Notably, the ordered pyrochlore structure is stable up to an A:B ratio of 2.75:1.25 or an equivalent $y$ value of 0.75 (in composition P3). In contrast, it is known that an order-disorder transition can occur at $y >$ ~0.1-0.3 in ternary $A_{2\pm y}B_{2\mp y}O_{7\mp y/2}$ oxides to form disordered defect-fluorite phase [31, 72]. This again suggests the unusual phase stability with ultrahigh-entropy compositions.

In order to further confirm the crystal structures (including the cubic-orthorhombic transition or PWT) and verify the elemental homogeneities at nanoscale, AC-STEM HAADF imaging and EDS elemental mapping have been carried out for three representative specimens, namely P3 (Fig. 5), W3 (Fig. 6), and MF (Fig. 7). For pyrochlore P3, the zone axis $[211]_p$ and two perpendicular planes $(1\bar{1}\bar{1})_p$ and $(01\bar{1})_p$ are marked in Fig. 5(a). For weberite W3, the zone axis $[110]_w$ and two perpendicular planes $(1\bar{1}0)_w$ and $(001)_w$ are indicated in Fig. 6(a). In the weberite specimen, lattice parameters $b_w$ and $c_w$ are too close (7.4323 Å vs. 7.4802 Å) to differentiate in STEM imaging. Nevertheless, digital image processing of the raw STEM image shows intensity modulations along the vertical direction and reveals cation ordering, which matches the atomic configuration of $(001)_w$ plane with the $[110]_w$ zone axis, but not the $(010)_w$ plane with the $[101]_w$ zone axis, as shown in Fig. 6(a). For fergusonite MF, the zone axis $[0\bar{1}\bar{1}]_{mf}$ and two planes, $(1\bar{1}1)_{mf}$ and $(100)_{mf}$, with an angle of ~86°, are labeled in Fig. 7(a). The fast Fourier transform (FFT) diffraction patterns of each STEM micrographs are also shown in Figs. 5(b), 6(b), and 7(b), respectively, to further validate the crystal structures and the crystallographic orientations.

In addition, STEM micrographs with the corresponding STEM-EDS elemental maps are shown in Fig. 5(c) for pyrochlore P3, Fig. 6(c) for weberite W3, and Fig. 7(c) for fergusonite MF. The combination of STEM HAADF imaging and EDS mapping have verified the homogeneous elemental distributions at atomic and nanometer scales, in addition to microscale as shown by SEM EDS (as shown in Supplementary Figs. S2-S10). In particular, STEM HAADF imaging and EDS mapping directly verified the presence of single ultrahigh-entropy pyrochlore phase in P3 and weberite phase in W3, just before and after the cubic-to-orthorhombic transition or the PWT, without detecting any nanoscale structural or compositional inhomogeneity (*i.e.*, no phase separation).

We further prepared and analyzed four specimens between the endmember fergusonite MF and the P0-W0 edge in the MF-P0-W0 compositional triangle shown in Fig. 2, which are labeled as P+MF0, P+MF1, P+MF2, and W+MF in Fig. 2. Dual-phase HECs have been observed for these four specimens (Supplementary Fig. S11-S14). By comparing their XRD spectra with the three endmembers (P0, W0, and MF) in Fig. 4, specimens P+MF0, P+MF1, and P+MF2 have been found to contain pyrochlore and ferguosnite dual phases. Similarly, specimen W+MF has been unveiled to comprise dual HEC phases of weberite and ferguosnite. Based on relative XRD peak intensities, SEM micrographs and EDS elemental maps, pyrochlore and weberite phases (of brighter contrast in SEM micrographs) are enriched in Sc (as well as Ti, Zr, and Hf for specimens P+MF0 to P+MF2). The ferguosnite phases (of darker contrast in SEM micrographs) are enriched in Nb (and probably Ta). Dual-phase HECs (of five metal cations) have been reported previously [51], while the four dual-phase HECs discovered here have 17, 19 or 21 different metals cations. We note that the formation of dual HEC phases is still significantly less than the maximum numbers of 17-21 equilibrium phases allowed by the Gibbs phase rule.

Fig. 8 shows the measured lattice parameters, Young's modulus, and thermal conductivity for the series of eight specimens with the compositions $(15RE_{1/15})_{2+x}(Ti_{1/4}Zr_{1/4}Ce_{1/4}Hf_{1/4})_{2-2x}(Nb_{1/2}Ta_{1/2})_xO_7$ ($0 \leq x \leq 1$), denoted as P0, P1, P2, P3, W3, W2, W1, and W0. The lattice parameters normalized to that of the primitive cubic fluorite cell ($a_f$) are shown in Fig. 8(a). In general, gradual increases in normalized lattice parameters have been observed from P0 ($x = 0$) to W0 ($x = 1$), which can be attributed to the increasing fractions of large RE elements [62]. However, the abrupt change in normalized lattice parameters due to the occurrence of PWT was observed between specimens P3 ($x = 0.75$) and W3 ($x = 0.8125$).

The measured Young's modulus ($E$) largely follows the rule-of-mixture (RoM) average of two endmembers P0 and W0, which decreases slightly from P0 to W0 (Fig. 8(b)). The measured thermal conductivity ($k$) at room temperature decreases, more appreciably, with increasing $x$, which also follows the RoM average of the endmembers P0 and W0 largely (Fig. 8(c)). However, there are small, but noticeable, changes in both measured Young's modulus and measured room-temperature thermal conductivity from P3 ($x = 0.75$) to W3 ($x = 0.8125$) with the occurrence of PWT (*i.e.* a cubic-orthorhombic transition), as shown in Fig. 8(b, c), which may be attributed to the changes of crystal symmetry and structure [32, 73]. The pyrochlore endmember P0 possesses an $E/k$ ratio of ~173 GPa·m·K·W$^{-1}$, on apar with those of high-entropy pyrochlore oxides [31, 39]. The weberite endmember W0 exhibits a high $E/k$ ratio of ~193 GPa·m·K·W$^{-1}$ (being thermally insulating yet stiff), consistent with ternary weberite niobates [73].

The temperature-dependent thermal conductivity was further measured for the specimens P3 ($x = 0.75$) and W3 ($x = 0.8125$), which represent the compositions just before and after the PWT. The measured thermal conductivity (after correction to remove the radiation contribution at high temperatures, following the model described in Supplementary Method) *vs.* temperature curves are shown in Fig. 9. The thermal diffusivity *vs.* temperature curves are shown in Supplementary Fig. S17. Despite a distinctive change in the crystal structure (from cubic to orthorhombic), the measured thermal conductivity *vs.* temperature curves for the pyrochlore P3 and weberite W3 are similar ("amorphous" behavior). The thermal conductivity monotonically increases slightly from ~1.27-1.28 W·m$^{-1}$·K$^{-1}$ at room temperature to ~1.41-1.42 W·m$^{-1}$·K$^{-1}$ at 600 °C. This increase is due to the rising specific heat capacity. Thermal diffusivity, as show in Supplementary Fig. S17, decreases from room temperature to 200 °C and then remains approximately a constant from 200 °C to 600 °C. The temperature dependences of the thermal conductivity and thermal diffusivity suggest diffusion-like behavior of the heat-carrying vibrational modes of P3 and W3 from 200 to 600 °C, where the thermal conductivity rises are mainly due to heat capacity contributions [74]. This behavior is similar to those observed for other high-entropy fluorite and pyrochlore oxides in a prior study [31]. Typically, lower symmetry means higher anharmonicity in a crystal, thereby causing stronger phonon Umklapp scattering and lower lattice thermal conductivity [75]. However, in this case, since the phonon scattering is so strong that the minimum thermal conductivity is already achieved in both materials, there is no difference in the thermal conductivity due to the different symmetries. Presumably, the cubic-to-orthorhombic transition (PWT) does not change the temperature-dependent thermal conductivity appreciably because the amorphous limit of phonon scattering may have already been achieved in the 21-component oxides at a relatively low temperature of 200 °C to produce low thermal conductivity. However, such phase transitions can alter other properties more appreciably.

## 4. Conclusions

We have investigated a 21-component oxide system to discover two new ultrahigh-entropy weberite and fergusonite phases and somewhat unexpected formation of ordered pyrochlore phases with ultrahigh-entropy compositions. While the Gibbs phase rule allows the formation of up to 21 equilibrium phases, nine out of the 13 compositions examined possess single HEC phases and the other four are dual-phase HECs. We further investigated a series of eight $(15RE_{1/15})_{2+x}(Ti_{1/4}Zr_{1/4}Ce_{1/4}Hf_{1/4})_{2-2x}(Nb_{1/2}Ta_{1/2})_xO_7$ ($0 \leq x \leq 1$) compositions, which all possess single-phase cubic pyrochlore or orthorhombic weberite structure, with an a fairly sharp pyrochlore-weberite (cubic-orthorhombic) transition occurring between $x = 0.75$ and 0.8125. This study reported nine new single-phase ultrahigh-entropy compositions in three crystal structures, each with 17 to 21 different metal cations and ultrahigh ideal configurational entropies. The ultrahigh-entropy compositions can exhibit phase stability deviating from the rule derived based on simple ternary oxides or conventional five-component HECs. For example, single-phase ordered pyrochlore structure (instead of disordered defect-fluorite structure) forms and is highly stable in 19-component and 21-component HECs with smaller $r_{A^{3+}}/r_{B^{4+}}$, large size disorder, and significant non-stoichiometries (*e.g.*, after adding significant amount of Nb$^{5+}$ and Ta$^{5+}$ cations). The thermal conductivity appears to already achieve the low amorphous limit above 200 °C, which does not change appreciable with the occurrence of a pyrochlore-weberite (cubic-orthorhombic) transition. In a broader context, this study has expanded the diversity and complexity of HECs, towards many-component CCCs and ultrahigh-entropy ceramics.

**Acknowledgement:** The work is supported by the National Science Foundation (NSF) in the Ceramics program via Grant No. DMR-2026193. This work utilized the shared facilities at the San Diego Nanotechnology Infrastructure of UCSD, a member of the National Nanotechnology Coordinated Infrastructure (supported by the NSF ECCS-1542148) and the Irvine Materials Research Institute (partially supported by NSF DMR-2011967 through UCI CCAM).

**Electronic Supplementary Material**

Supplementary material is available in the online version of this article at https://doi.org/..., including

- Supplementary Method (Thermal Conductivity Measurements and Corrections)
- Supplementary Tables S1 and S2
- Supplementary Figures S1-S17.

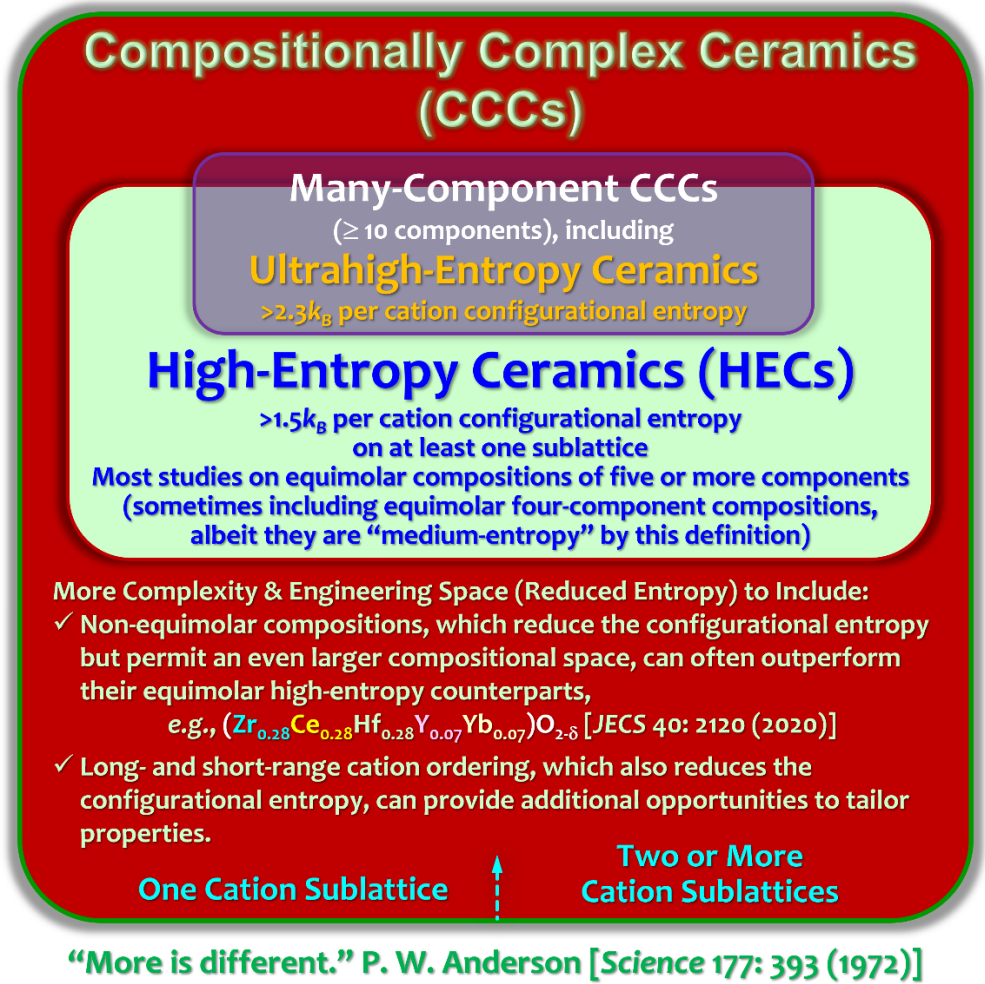

**Fig. 1.** Schematic illustrations of high-entropy ceramics (HECs) and compositionally complex ceramics (CCCs), along with the proposed many-component CCCs and ultrahigh-entropy ceramics. Following a convention commonly used in high-entropy alloys [50], we can define HECs to refer to ceramics possessing >1.5$k_B$ per cation ideal configurational entropy on at least one sublattice [1], where $k_B$ is the Boltzmann constant. Most prior studies of HECs focused on equimolar compositions of five or more components, but some studies also included equimolar four-component compositions, albeit they are "medium-entropy" by this definition [1]. We also proposed to broaden HECs to CCCs to include medium-entropy and/or non-equimolar compositions and consider long- and short-range cation ordering, which reduce the configurational entropy, but provide additional opportunities to tailor properties [1]. Many-component CCCs (≥10 components) offer further compositional complexity (and tunability). Moreover, we can define ultrahigh-entropy ceramics (a subclass of HECs and a subclass of many-component CCCs) as those HECs with >2.3$k_B$ per cation ideal configurational entropy on at least one sublattice; noting that ln(10) ≈ 2.3, so that equimolar many-component CCCs with ≥10 disordered (randomly dissolved) components should be ultrahigh-entropy. We note that these definition thresholds adopted here (5 *vs.* 10 components, and 1.5$k_B$ *vs.* 2.3$k_B$ per cation) are somewhat arbitrary and subjective. Like HECs *vs.* CCCs, non-equimolar compositions and cation ordering can reduce the configurational entropy of many-component CCCs so that they (may or) may not be ultrahigh-entropy (or even high-entropy) ceramics. Two or multiple cation sublattices, anion mixing, mixed ionic-covalent-metallic bonding, and defects (*e.g.*, oxygen vacancies and aliovalent doping) can further increase the complexity of CCCs. It is important to note that maximizing the entropy may not always to be beneficial or necessary; the complexities in (many-component) CCCs offer more opportunities to tailor properties, where different or new phenomena can emergent in complex systems, because "more is different" [52].

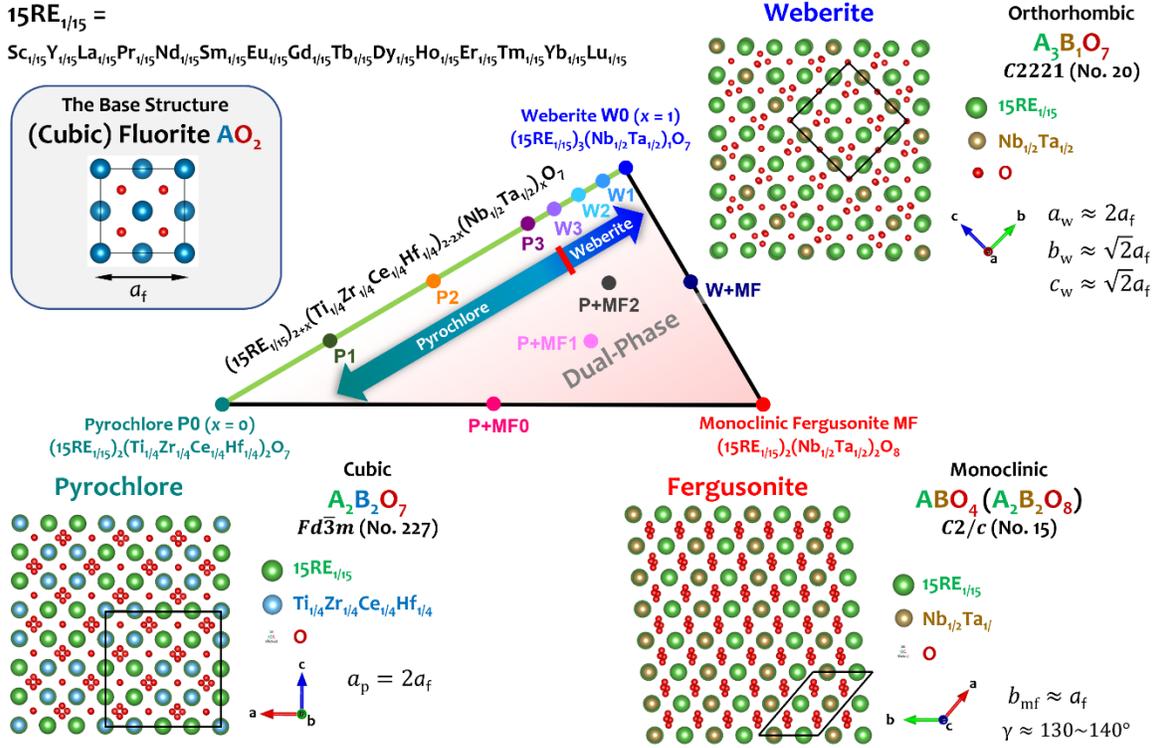

**Fig. 2.** Schematic illustration of the P0-W0-MF compositional triangle investigated in this study. The three endmembers are single-phase high-entropy ceramics (HECs) that form three different fluorite-based crystal structures. First, endmember P0, $(15RE_{1/15})_2(Ti_{1/4}Zr_{1/4}Ce_{1/4}Hf_{1/4})_2O_7$, where $15RE_{1/15}$ refers to the equimolar mixture of 15 rare earth (RE) elements (labeled on the top of the figure), is in the cubic $A_2B_2O_7$ pyrochlore structure ($Fd\bar{3}m$, No. 227). Second, endmember W0, $(15RE_{1/15})_3(Nb_{1/2}Ta_{1/2})_1O_7$, is in the $Y_3TaO_7$-prototyped orthorhombic $A_3BO_7$ weberite-type structure ($C222_1$, No. 20). Third, endmember MF, $(15RE_{1/15})_2(Nb_{1/2}Ta_{1/2})_2O_8$, is the $YNbO_4$-prototyped $ABO_4$ (or $A_2B_2O_8$) monoclinic fergusonite-type structure ($C2/c$, No. 15). We selected 13 compositions in this P0-W0-MF compositional triangle to investigate in this study. These include a series of eight compositions on the P0-W0 edge (the light green line) of $(15RE_{1/15})_{2+x}(Ti_{1/4}Zr_{1/4}Ce_{1/4}Hf_{1/4})_{2-2x}(Nb_{1/2}Ta_{1/2})_xO_7$ ($0 \leq x \leq 1$), which form single-phase pyrochlore or weberite structure (with a rather abrupt transition). These eight compositions denoted as P0, P1, P2, P3, W3, W2, W1, and W0. In addition, we examined the other endmember $(15RE_{1/15})_2(Nb_{1/2}Ta_{1/2})_2O_8$ that is in single-phase monoclinic fergusonite (donated as "MF"), and we further observed three pyrochlore-MF dual-phase specimens denoted as P+MF0, P+MF1, and P+MF2, as well as a weberite-MF dual-phase specimen denoted as W+MF. Since all these three structures (pyrochlore, weberite, and fergusonite) can be considered as derivative structures based on a primitive cubic fluorite structure ($Fm\bar{3}m$, No. 225), their unit cells are delineated and the lattice parameter relationships between these three structures and the primitive fluorite cell are also noted in the figure. Here, we refer the monoclinic fergusonite structure as "MF" and denote their lattice parameters as $a_{mf}$, $b_{mf}$, and $c_{mf}$ to avoid confusion with the cubic fluorite structure ($a_f$).

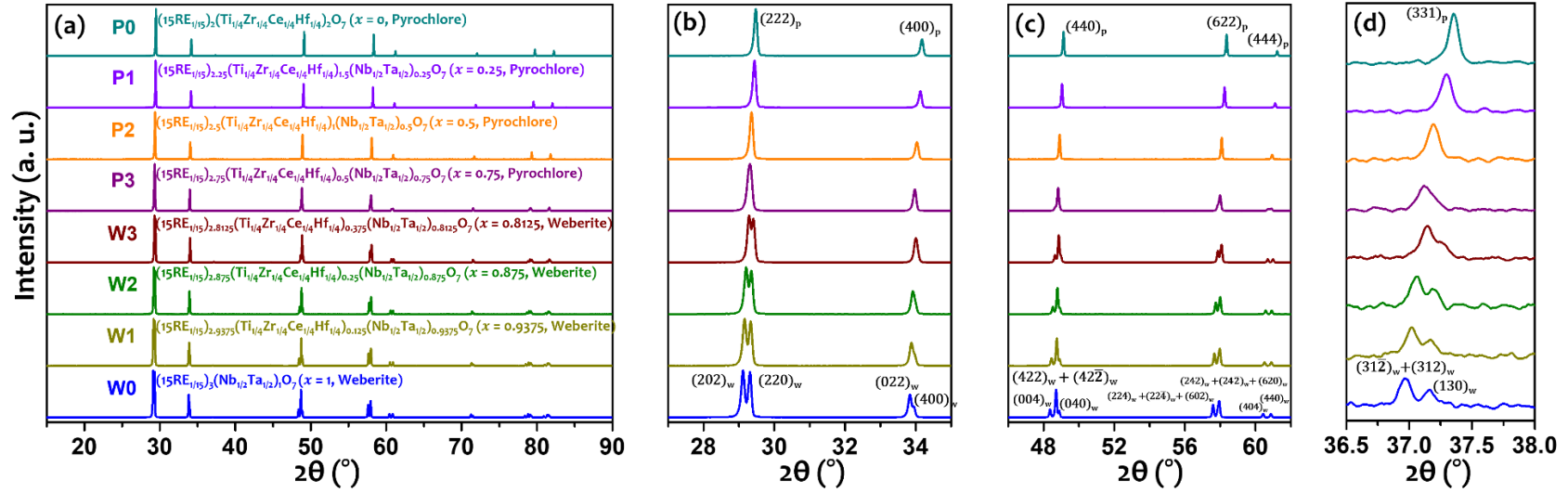

**Fig. 3.** (**a**) XRD patterns of eight specimens in the Pyrochlore-Weberite series of P0, P1, P2, P3, W3, W2, W1, and W0 with compositions of $(15RE_{1/15})_{2+x}(Ti_{1/4}Zr_{1/4}Ce_{1/4}Hf_{1/4})_{2-2x}(Nb_{1/2}Ta_{1/2})_xO_7$ ($x$ = 0, 0.25, 0.5, 0.75, 0.8125, 0.875, 0.9375, and 1). (**b-d**) Enlarged views of the peak evolutions from cubic pyrochlore to orthorhombic weberite. Due to the orthorhombic structure of weberite, where $b_w \approx \sqrt{2}a_f$, $c_w \approx \sqrt{2}a_f$, but $b_w \neq c_w$, its XRD reflections usually show double peaks. The $(331)_p$ in Panel (d) is the pyrochlore superstructure peak (over the primitive cubic fluorite). Detailed XRD and SEM EDS elemental mapping results of these eight compositions are documented in Supplementary Fig. S2-S9. Note that the orthorhombic weberite structure of lower symmetry shows more XRD diffraction peaks.

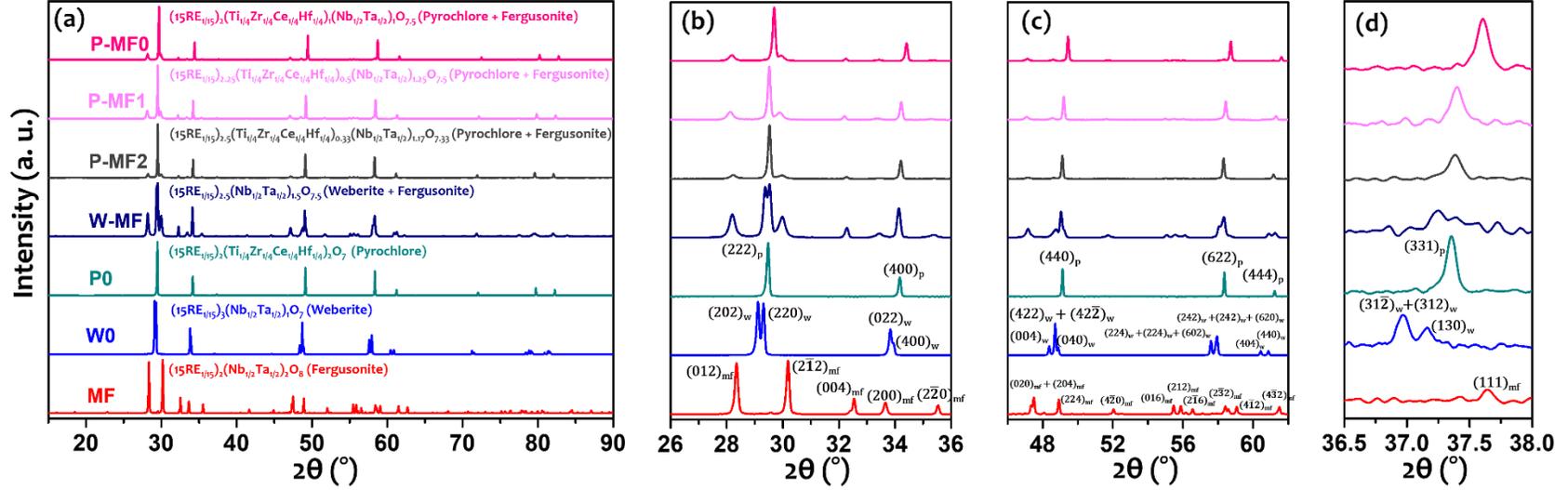

**Fig. 4.** (**a**) XRD patterns of four dual-phase specimens, along with monoclinic fergusonite MF $(15RE_{1/15})_2(Nb_{1/2}Ta_{1/2})_2O_8$. XRD patterns of the pyrochlore P0 $(15RE_{1/15})_2(Ti_{1/4}Zr_{1/4}Ce_{1/4}Hf_{1/4})_2O_7$ and weberite W0 $(15RE_{1/15})_3(Nb_{1/2}Ta_{1/2})_1O_7$, are also juxtaposed for comparison to ratify the dual phases of the other four compositions (**b-d**) Enlarged views of the detailed peak comparisons. Detailed XRD and SEM EDS elemental mapping results of all these seven compositions are documented in Supplementary Fig. S2 and S9-S14. In single-phase specimens, the orthorhombic weberite W0 and monoclinic fergusonite MF possess more XRD diffraction peaks (compared with cubic pyrochlore P0) due to their low symmetries.

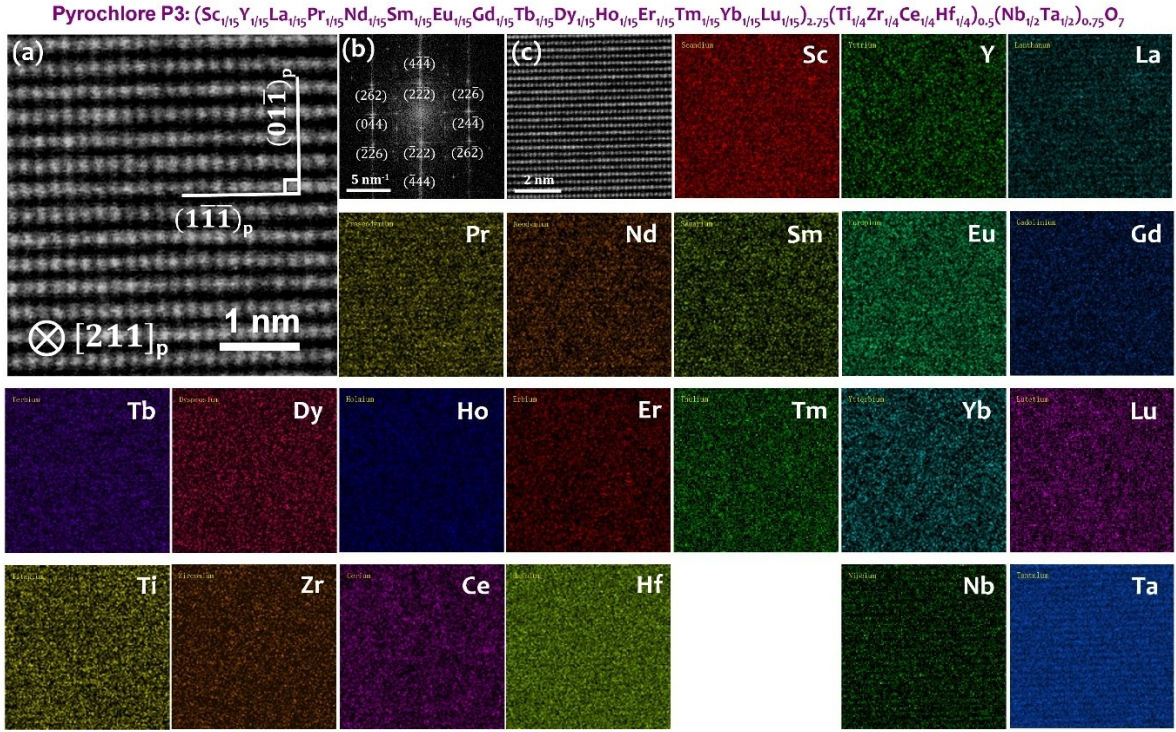

**Fig. 5.** (**a**) STEM HAADF image of P3 $(15RE_{1/15})_{2.75}(Ti_{1/4}Zr_{1/4}Ce_{1/4}Hf_{1/4})_{0.5}(Nb_{1/2}Ta_{1/2})_{0.75}O_7$, which illustrates the pyrochlore atomic structure. The zone axis $[211]_p$ and two perpendicular planes $(1\bar{1}\bar{1})_p$ and $(01\bar{1})_p$ of the pyrochlore structure are marked. (**b**) The FFT diffraction pattern, which further validates the aforementioned crystallographic orientations. (**c**) STEM micrograph with the corresponding EDS elemental maps, demonstrating the homogenous elemental distributions at the nanoscale.

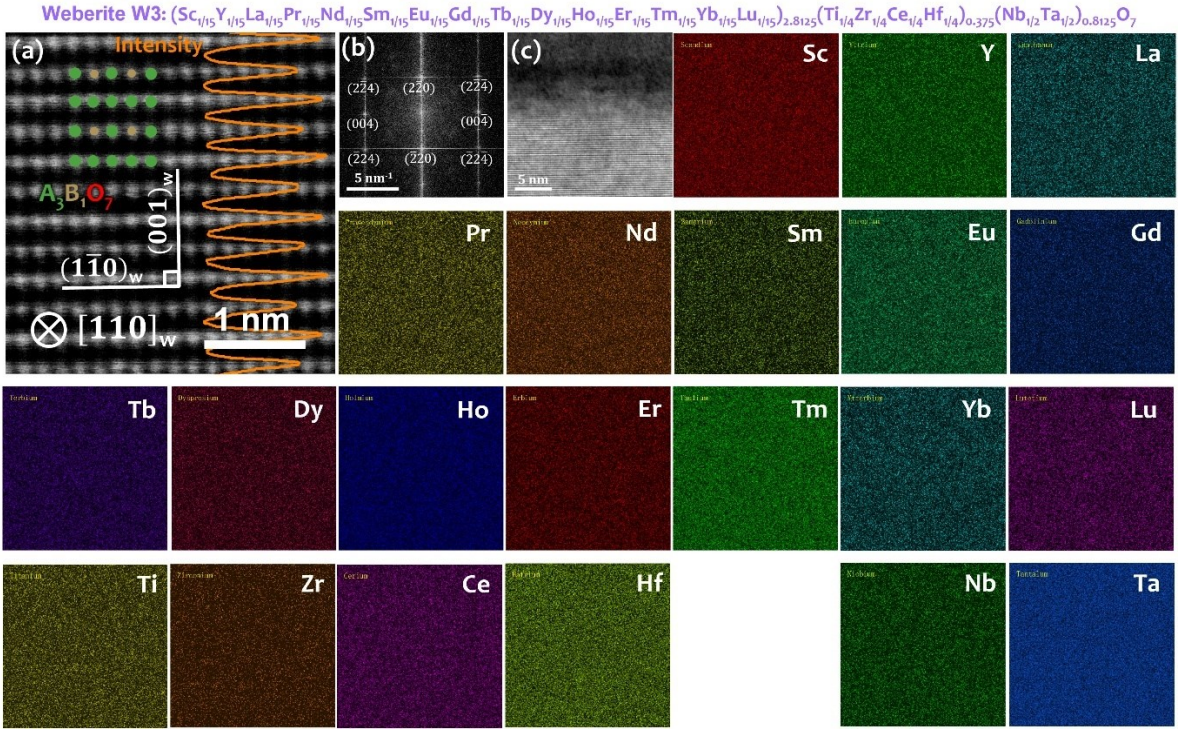

**Fig. 6.** (**a**) STEM HAADF image of W3 $(15RE_{1/15})_{2.8125}(Ti_{1/4}Zr_{1/4}Ce_{1/4}Hf_{1/4})_{0.375}(Nb_{1/2}Ta_{1/2})_{0.8125}O_7$, which illustrates the weberite-type atomic structure. The zone axis $[110]_w$ and two perpendicular planes $(1\bar{1}0)_w$ and $(001)_w$ of the weberite structure are indicated. In this weberite specimen, lattice parameters $b_w$ and $c_w$ are too close (7.4323 Å *vs.* 7.4802 Å). Nevertheless, digital image processing of the raw STEM-HAADF image shows intensity modulations along the vertical direction to reveal cation ordering, which matches the atomic configuration of $(001)_w$ plane with the $[110]_w$ zone axis, but not the $(010)_w$ plane with the $[101]_w$ zone axis. (**b**) The FFT diffraction pattern, which further validates the aforementioned crystallographic orientations. (**c**) STEM micrograph with the corresponding EDS elemental maps, demonstrating the homogenous elemental distributions at the nanoscale.

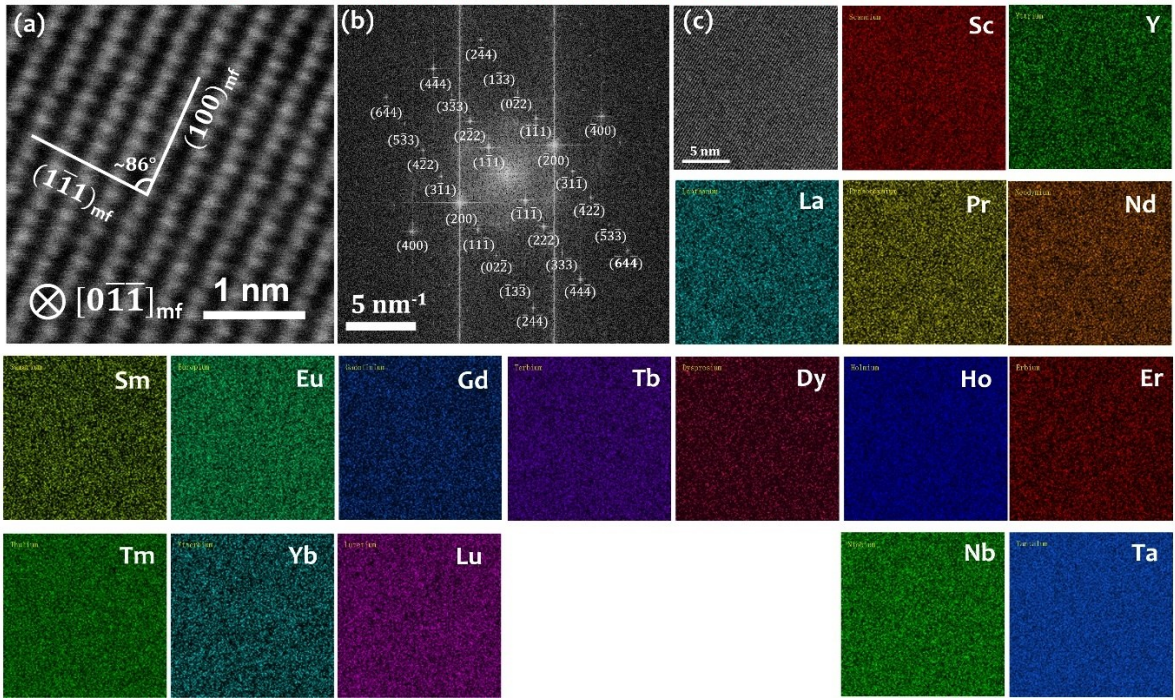

**Fig. 7.** (**a**) STEM HAADF image of MF $(15RE_{1/15})_2(Nb_{1/2}Ta_{1/2})_2O_8$, which illustrates the monoclinic fergusonite-type atomic structure. The zone axis $[0\bar{1}\bar{1}]_{mf}$ and two planes, $(1\bar{1}1)_{mf}$ and $(100)_{mf}$ of the fergusonite structure, which are at an angle of ~86°, are labeled accordingly. (**b**) The FFT diffraction pattern, which further validates the aforementioned crystallographic orientations. (**c**) STEM micrograph with the corresponding EDS elemental maps, demonstrating the homogenous elemental distributions at the nanoscale.

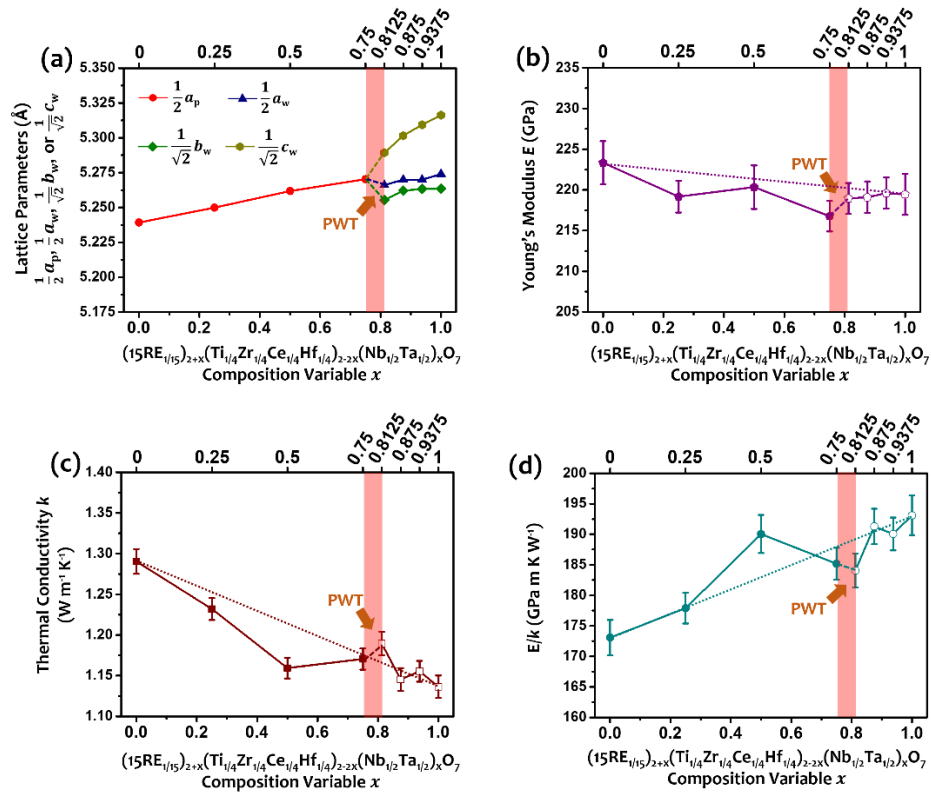

**Fig. 8.** (**a**) Lattice parameters and room-temperature (**b**) Young's modulus ($E$), (**c**) thermal conductivity ($k$), and (**d**) $E/k$ ratio measured for the eight specimens in the Pyrochlore-Weberite series of the general compositional formula $(15RE_{1/15})_{2+x}(Ti_{1/4}Zr_{1/4}Ce_{1/4}Hf_{1/4})_{2-2x}(Nb_{1/2}Ta_{1/2})_xO_7$ with composition variable $0 \leq x \leq 1$. Arrows and dashed lines are used to denote the pyrochlore-to-weberite transition (PWT). Dotted lines represent the rule-of-mixture averages from two endmembers. The lattice parameters are obtained from unit cell refinements of the XRD spectra and are normalized to the primitive fluorite cell ($a_f$) for direct comparison.

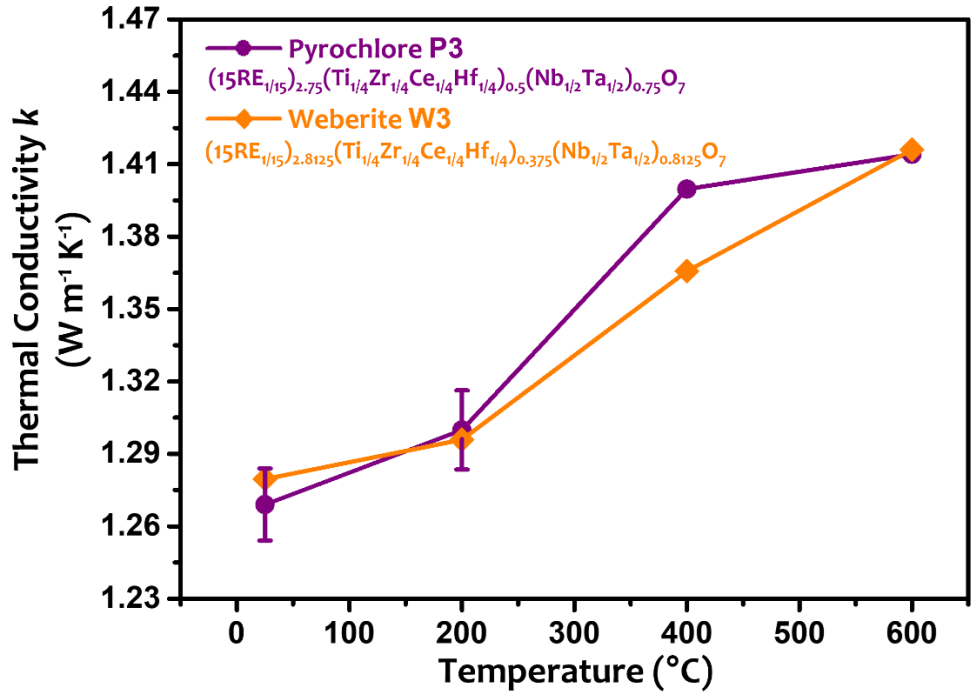

**Fig. 9.** Measured thermal conductivity (after removing the radiation contribution and corrected for porosity) *vs.* temperature curves of pyrochlore P3 $(15RE_{1/15})_{2.75}(Ti_{1/4}Zr_{1/4}Ce_{1/4}Hf_{1/4})_{0.5}(Nb_{1/2}Ta_{1/2})_{0.75}O_7$ (denoted by purple circles) and weberite W3 $(15RE_{1/15})_{2.8125}(Ti_{1/4}Zr_{1/4}Ce_{1/4}Hf_{1/4})_{0.375}(Nb_{1/2}Ta_{1/2})_{0.8125}O_7$ (denoted by orange diamond), which are the two compositions just before and after the occurrence of the PWT. Notably, the *k* values reported here represent the true (net) thermal conductivities after removing the radiative heat transfer contributions at high temperatures using a transparent model (differing from the apparent thermal conductivities calculated using the standard model from measured thermal diffusivities). The procedure to remove the radiation contribution is described and discussed in the Supplementary Method.